\documentclass[aps,pra,twocolumn,groupedaddress]{revtex4-1}
\usepackage{nicefrac}
\usepackage{amssymb}
\usepackage{amsmath}
\usepackage{bbold}
\usepackage{graphicx}
\usepackage{hyperref}

\newcommand{\ket}[1]{\left\vert #1 \right\rangle}
\newcommand{\bra}[1]{\left\langle #1 \right\vert}
\begin{document}


\title{Study of Multiple Rounds of Error Correction in Solid State NMR QIP}


\author{Ben Criger}
\affiliation{Institute for Quantum Computing}
\affiliation{Department of Physics and Astronomy, University of Waterloo}
\email[dcriger@iqc.ca]{}
\author{Osama Moussa}
\affiliation{Institute for Quantum Computing}
\affiliation{Department of Physics and Astronomy, University of Waterloo}
\author{Raymond Laflamme}
\affiliation{Institute for Quantum Computing, University of Waterloo}
\affiliation{Perimeter Institute for Theoretical Physics, Waterloo, Ontario}
\affiliation{Department of Physics and Astronomy, University of Waterloo}


\date{\today}

\begin{abstract}
Methods to control errors will be essential for quantum information processing. It is widely believed that fault-tolerant quantum error correction is the leading contender to achieve this goal.  Although the theory of fault-tolerant quantum error correction is very well understood, experimental implementation has been lagging.  We study the feasibility of implementing repeated rounds of quantum error correction with refreshed ancillas in solid state nuclear magnetic resonance (NMR). In particular we study the procedure proposed for extracting entropy that consists of two stages; an error correcting code optimized to function at finite temperature, and an implementation of heat-bath algorithmic cooling to refresh the ancilla qubits. Two algorithms are presented which implement this method, one for performing tests on 4 qubits, the other for practical implementation on 6 qubits. The effects of imperfect implementation are examined in both the error correction and refreshing stages. 
\end{abstract}
\pacs{03.67.Pp}
\keywords{Heat Bath Algorithmic Cooling, Optimized Error Correction, Nuclear Magnetic Resonance, Quantum Information, Quantum Computing}

\maketitle

\section{Introduction}
It is possible to encode information onto systems which are governed by the laws of quantum mechanics. The extra computational freedom granted by such an approach can enable exponentially faster algorithms for a growing set of tasks. Included among these are the factoring of large numbers \cite{shorfactoring}, database search \cite{groversearch}, and simulation of physical systems governed by quantum mechanics \cite{springerlink:10.1007/BF02650179}.

A challenge to the implementation of quantum algorithms in physical systems, with which this manuscript is concerned, is the process of decoherence. Quantum states are inherently delicate; processes beyond experimental control drive quantum systems toward classical behaviour. This presents a challenge to the implementation of quantum computing, whose efficacy depends on the presence of non-classical behaviour. 

\begin{figure}[h!]
\centering
\includegraphics[width=0.25\textwidth]{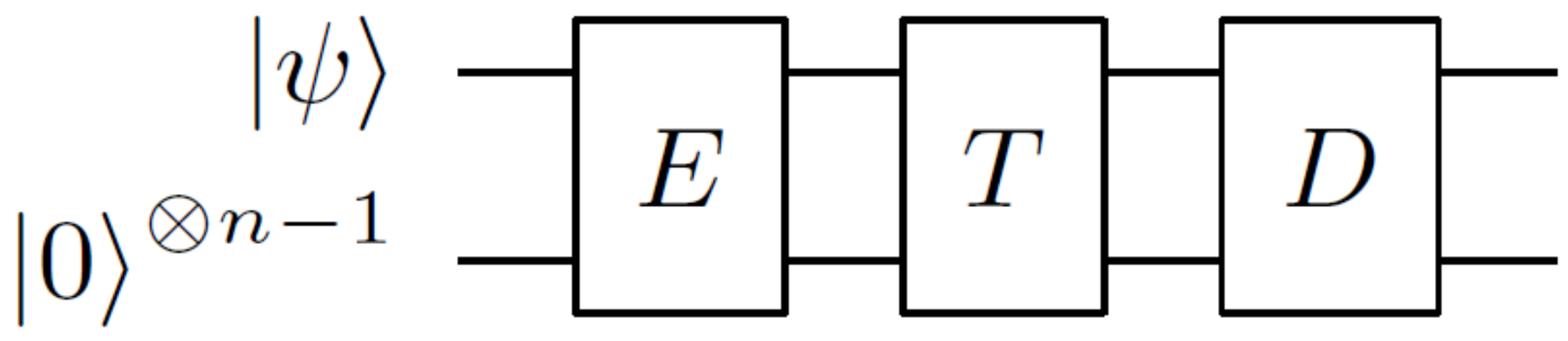}
\caption{Sketch of a generic error correction procedure, decomposed into three phases. First, the one-qubit basis states (typically labelled $\ket{0}$ and $\ket{1}$) are \textit{encoded} ($E$) into two distinct $n$-qubit states. Then the encoded state is \textit{transmitted} ($T$) through a noisy channel, or stored in an unreliable medium. In the third phase, the state is \textit{decoded} ($D$), providing an output state with improved fidelity to the original state. The ancilla qubits, incorporated in the encoding phase, are then disposed of.}
\label{fig:generic}
\end{figure}

There are two principal means of reducing or eliminating these errors. The first is to further isolate the quantum computer from the external environment, reducing the unwanted interactions directly. This is the tactic employed by  topological quantum computing \cite{RevModPhys.80.1083}, which uses topologically-protected states to store information. The second method, discussed in this manuscript, is the use of error correction protocols (as seen in Figure \ref{fig:generic}). These are based on encoding a one-qubit state into multiple qubits, allowing an error process to act upon the encoded state, and decoding/correcting the original state.

Some elements of fault-tolerant quantum error correction have already been implemented.  In many systems, phase errors are dominant and thus the 3 qubit phase correcting code (\cite{PhysRevLett.81.2152}, \cite{iontraps}, \cite{PhysRevB.77.214528}, \cite{PhysRevA.82.012319}) and a 5-bit code \cite{PhysRevLett.86.5811}
have been implemented to reduce the noise. 

In the case of NMR, it is not possible to make projective measurement adding difficulty to the task of refreshing qubits for more rounds of error correction. It is however possible to use algorithmic cooling as suggested by (\cite{Umesh99molecularscale},\cite{PhysRevLett.94.120501}) and first demonstrated in \cite{PhysRevLett.100.140501}.
More recently, another building block for fault tolerance was demonstrated: magic state distillation (see \cite{PhysRevA.71.022316} for theoretical background, \cite{hubbardmastersthesis} for details of implementation).

In order to practically implement quantum computing, it is necessary to have the ability to perform multiple iterations of error correction, so that states can be preserved against arbitrary noise over arbitrary time scales. This arises from the presence of second-order error terms.  

A common assumption underlying the design of error correcting codes is that ancilla qubits (qubits employed in encoding in addition to the qubit on which the message is stored) are highly polarized, typically initialized in the state $\ket{0}$. These ancilla qubits, in practice, are in short supply and need to be `refreshed' to high polarization in order to run multiple iterations of error correction.

In section II of this manuscript, the proposed experimental setup is discussed, in order to provide context to the algorithm detailed in the following sections. In section III of this manuscript, a modified version of the traditional three-bit error correcting code is presented, with the experimental implementation of multiple rounds of error correction as the eventual goal. In Section IV, a known mechanism for transferring entropy is analysed, and the effects of imperfect control in experimental error correction are discussed. 
\section{Experimental Setup}
In order to provide context for the code presented, we first detail the experimental situation for which this code is designed. This does not preclude the code presented in later sections from being used in other quantum computing architectures, provided that similar resources are present.

Single-crystal solid-state NMR employs a large ensemble of identical molecules with fixed orientation on a crystal lattice. Each nucleus in a molecule precesses about an axis defined by the large external magnetic field at its Larmor frequency $\omega $, which is the product of a `chemical shift' term $ (1+\delta)$ which arises from local electrons screening the external magnetic field, the gyromagnetic ratio of the nucleus in question $\gamma$, and the magnetic field $ B_0$. The state of the spins is measured by analysing the response of the system to an rf pulse. Two-qubit operations are obtained through dipole-dipole coupling. These, together, allow for universal control, and the large number of identical quantum information processors produces a detectable signal. The advantages of this implementation of quantum computing are described further in \cite{Gershenfeld17011997}, \cite{CLK00}, \cite{BMR06}.  

\begin{figure}[h!]
\centering
\includegraphics[width=0.45\textwidth]{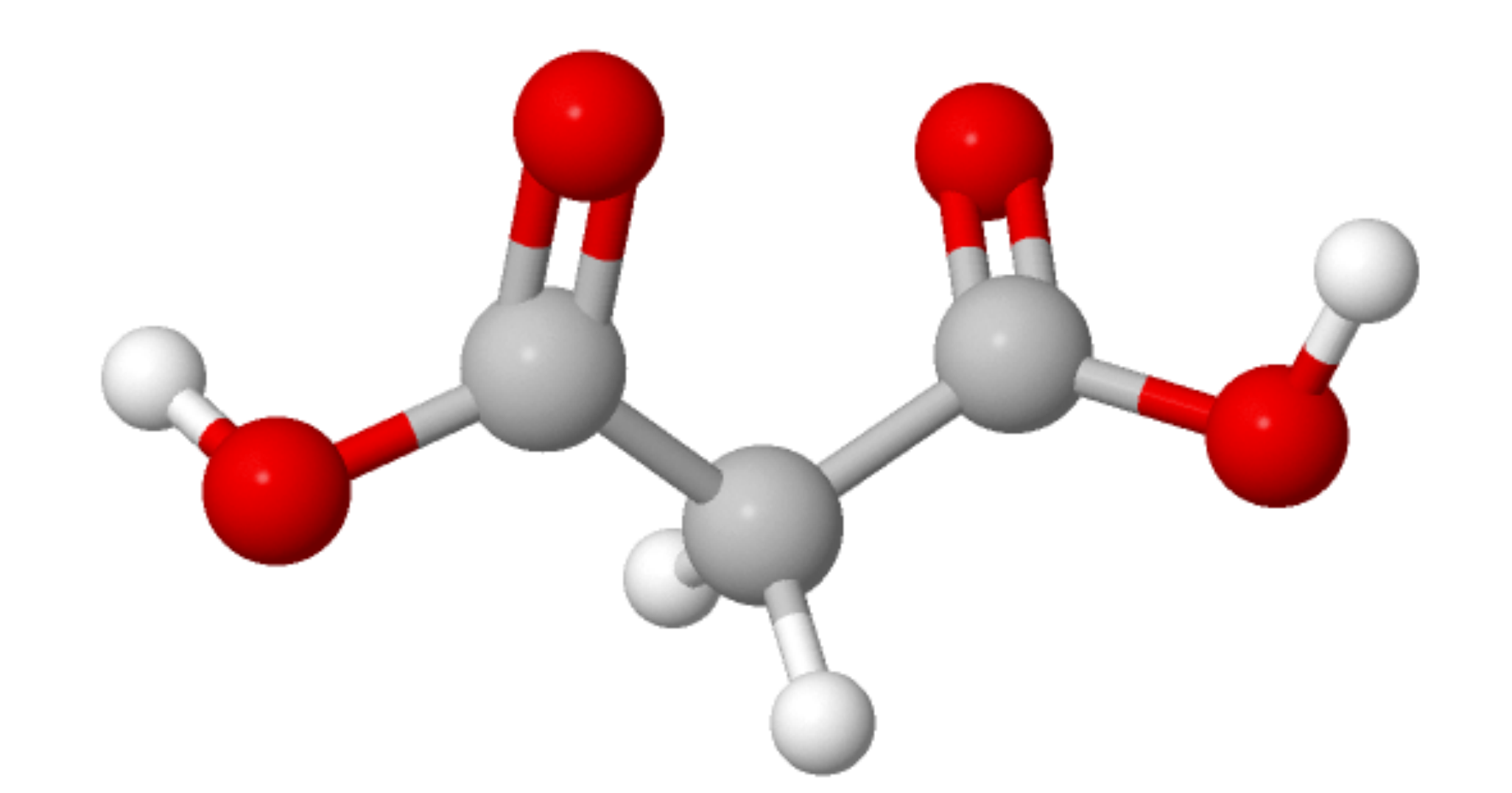}
\caption{Malonic acid. This molecule encodes 3 qubits in NMR experiments, and is the molecule used in \cite{PhysRevLett.100.140501}. A malonyl radical can be formed by removing one of the protons; the remaining electron can be coupled to the quantum system of interest. }
\label{fig:malonic}
\end{figure}

The physical system employed in \cite{CLK00}, \cite{BMR06} and \cite{PhysRevLett.100.140501} merits discussion, since it is similar in many respects to the system envisioned for the experimental implementation of the procedures discussed below. The architecture used is single-crystal solid-state nuclear magnetic resonance in which malonic acid (the structure of which is shown in Figure \ref{fig:malonic}) with three $^{13}$C is present in the crystal sample at $\sim3\%$ abundance. The remainder of the molecules have natural abundance $^{13}$C. These unlabelled molecules prvent unwanted interactions between labelled molecules by ensuring that the probability of 2 triply-labelled molecules neighbouring one another in the crystal lattice remains negligibly small.  
The algorithms presented in the sections below will require 4 and 6 qubits, and a high-polarization bath. Larger molecules will be required in order to perform these experiments. 

\begin{figure}[h!]
\centering
\includegraphics[width=0.35\textwidth]{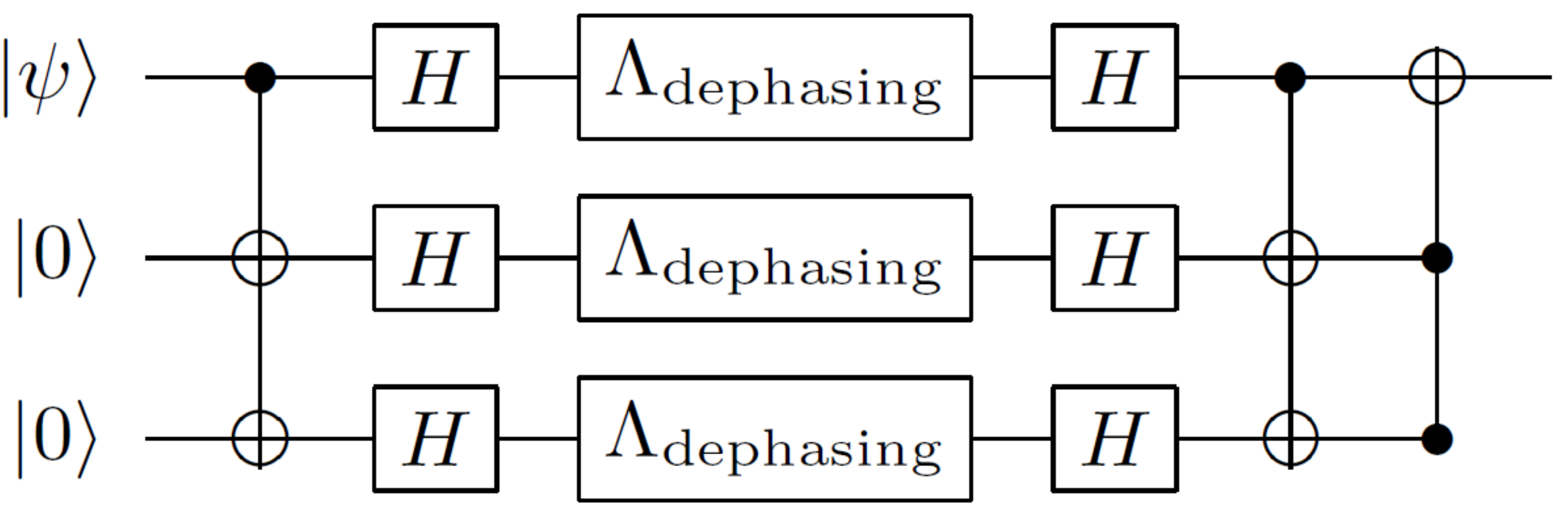}
\caption{The 3-bit error-correcting code, modified to protect against dephasing errors. The first qubit is prepared in an arbitrary state $\ket{\psi}=\alpha \ket{0} + \beta \ket{1}$. The other two qubits are prepared in the state $\ket{0}$. The state on the first qubit is `encoded' to $\alpha \ket{000} + \beta \ket{111}$. Then, each qubit is subjected to the Kraus map defined in Equation \ref{eq:phaseflipkraus}. The most likely outcomes ($\mathcal{T}^{\otimes 3}$ has no effect, or flips a single qubit) map the 3-qubit codeword into two orthogonal subspaces; $\ket{000} \rightarrow \left \lbrace \ket{000},\,\ket{001},\,\ket{010},\,\ket{100} \right \rbrace $ and $\ket{111} \rightarrow \left \lbrace \ket{111},\,\ket{110},\,\ket{101},\,\ket{011} \right \rbrace $. These errors can then be distinguished and corrected, to yield an error probability that scales as $p^2$ instead of $p$, since two independent errors must occur in order to induce an error in the decoded state.}
\label{fig:shorcode}
\end{figure}

In NMR quantum computing, the dominant source of error is `dephasing'. Any process that involves uncontrolled rotations about the quantization axis (the $z$-axis) results in a dephasing error. In NMR, this uncontrolled rotation is either the result of a difference in the magnetic field to which each individual spin is subjected, or the result of parasitic coupling between qubits and non-qubit particles present in the crystal. We represent this as a quantum channel $\Lambda_{\textrm{dephasing}}$ involving the random application of the Pauli $\hat Z$ operator with probability $p$. This channel can be expressed in the operator-sum representation:
\begin{equation}
\Lambda_{\textrm{dephasing}} = \left \lbrace \sqrt{1-p} \hat 1, \, \sqrt{p} \hat Z \right \rbrace\\
\label{eq:phaseflipkraus}
\end{equation}
\begin{align}
\Lambda(\rho) &= \sum_j \Lambda_j \rho \Lambda_j^{\dagger} \nonumber \\
& = (1-p) \rho + p\hat Z \rho \hat Z
\label{eq:sumoperator}
\end{align}
Here, $\hat 1$ is the identity operator. We can correct such errors with a variation on the traditional 3-bit error correcting code (\cite{PhysRevA.52.R2493}, \cite{braunstein-1996}), that translates dephasing errors into `bit-flip' errors, shown in Figure \ref{fig:shorcode}.
This error correcting code employs Hadamard gates to transform the dephasing channel into the more familiar bit-flip channel. The Hadamard gate transforms the states $ \lbrace \ket{0}, \, \ket{1} \rbrace$ to equal superpositions of $\ket{0}$ and $\ket{1}$, its matrix form is 
\begin{equation}
\hat H=\frac{1}{\sqrt{2}}\left[ \begin{array}{cc} 1& 1 \\ 1 & -1\end{array}\right].
\end{equation}
In order to determine the effect of the process $\hat H \Lambda_{\textrm{dephasing}} \hat H$, we note two properties of the Hadamard operator:
\begin{equation}
\hat H^2=\hat 1; \,\, \hat H \hat Z \hat H = \hat X 
\end{equation}
\begin{flalign}
\therefore \hat H \Lambda_{\textrm{dephasing}} \hat H &= \left \lbrace \sqrt{1-p}\hat 1,\,\, \sqrt{p} \hat X \right \rbrace  \nonumber \\
 & = \Lambda_{\textrm{bit-flip}}.
\end{flalign}

Another important feature of NMR quantum computing is the ability to form radicals (molecules with one proton missing) with electron spins that can be controlled and coupled to the nuclear system through ENDOR \cite{spindynamics}. If electron-nucleus interactions can be controlled, one or more electrons in each molecule can be used as a heat bath. Even though all components of a given molecule are at the same temperature, the polarization in the spin degree of freedom of particles inside a molecule may differ, since the polarization depends also on the magnetic field, and most importantly on the gyromagnetic ratio $\gamma$ of the particle in question:
\begin{equation}
\epsilon = \tanh \left( \frac{\hbar \gamma B_0}{k_BT}\right)
\end{equation}
Since the gyromagnetic ratio of the electron is $\sim 10^3$ times greater than the gyromagnetic ratio of $^{13}$C, the electron is more polarized than any nucleus at thermal equilibrium. A transfer of polarization from electrons to nuclei is, effectively, an exchange of entropy or `heat' between the two particles. This capacity for entropy exchange will be used to polarize $^{13}$C beyond the bath polarization. At temperatures on the order of 1 K, the electron polarization is of order 1, allowing for unit polarization on the nucleus as well. 
\section{Optimal Error Correction}
\begin{figure}[h!]
\centering
\includegraphics[width=0.4\textwidth]{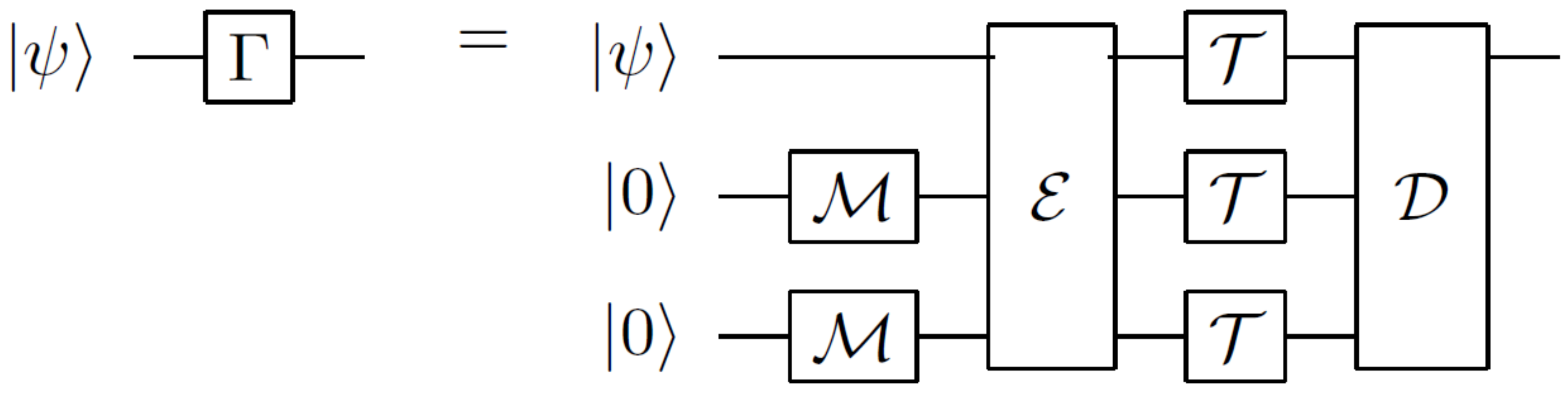}
\caption{A generic 3-qubit error correction procedure, with the encoding and decoding procedures replaced by unknown processes. The presence of initially mixed ancilla qubits is modelled with an additional channel before the encoding operation.}
\label{fig:modelchannel}
\end{figure}

The existing three-qubit code has been developed with the underlying assumption of polarized ancillary qubits which can be quickly exchanged. This code is presented in Figure \ref{fig:shorcode}. For comparison, the error channels whose effects must be corrected are presented in Figure \ref{fig:modelchannel}. Note that, in addition to the channel which is corrected by the traditional code, there is a second channel which maps the ideal (polarized) ancilla qubits to thermal states. In this section of the manuscript, we review the methods used to optimize error-correcting codes for experimental implementation. This section concludes with the presentation of an improved error correcting code analysed throughout the remainder of the manuscript and limits on the degree of mixture that is tolerable for such a code to be useful.  

\subsection{Background}
In order to optimize error-correcting codes, we require an objective function to quantify their performance; a measure of the similarity between the input and output states. The `channel fidelity' $F_C$, a special case of Schumacher's entanglement fidelity \cite{PhysRevA.54.2614} suits this purpose:
\begin{equation}
F_C(\Lambda)=\bra{\Omega} (\Lambda \otimes \hat 1)\left[ \ket{\Omega}\bra{\Omega} \right] \ket{\Omega} = \dfrac{1}{4^n}\sum_j \vert \textrm{tr} (\Lambda_j)\vert^2 ; 
\label{eq:channelfidelity}
\end{equation}
\vspace{-12pt}
\begin{equation*}
\textrm{where }\ket{\Omega} = \frac{1}{\sqrt{2^n}} \sum_{j \in \lbrace 0,1 \rbrace ^n}\ket{jj}
\end{equation*}
Here, $\Lambda$ is a quantum channel representing all of the error-correction operations (encoding, error, decoding), with Kraus operators $\Lambda_j$. $\ket{\Omega}$ is an entangled state with $2n$ qubits, where the error correction is to take an $n$-qubit state as the input. By extending the quantum system, the channel fidelity provides a concise means of averaging fidelity over all $n$-qubit states. 

With the metric in Equation \ref{eq:channelfidelity}, we can now define a criterion for useful error correction. We require that $F_C(\mathcal{D} \circ \mathcal{T} \circ \mathcal{E} \circ \mathcal{M} \circ \mathcal{A})$(where $\mathcal{D} \circ \mathcal{T} \circ \mathcal{E} \circ \mathcal{M} \circ \mathcal{A}$ is a composite process consisting of appending ancilla qubits in the state $\ket{00}\bra{00}$, $\mathcal{A}[\rho] = (\hat 1 \otimes \ket{00}\bra{00})\rho(\hat 1 \otimes \ket{00}\bra{00})$, mixing the states on the ancilla ($\mathcal{M}$, see Equation \ref{eq:mixingchannel}), encoding the resulting 3-qubit state ($\mathcal{E}$), transmission ($\mathcal{T}$) and decoding ($\mathcal{D}$)) is greater than $F_C(\mathcal{T})$. $F_C(\mathcal{T})$ is trivial to evaluate:
\begin{equation}
\mathcal{T}=\Lambda_{\textrm{dephasing}} = \left \lbrace \sqrt{1-p}\hat 1,\,\,\sqrt{p}\hat Z \right \rbrace \\
\end{equation}
\begin{flalign}
F_C(\mathcal{T}) &= \dfrac{1}{4}\sum_j \vert \textrm{tr} (\mathcal{T}_j)\vert^2 \nonumber\\
&= \dfrac{1}{4} \left((1-p) \vert \textrm{tr} (\hat 1) \vert^2 +p \vert \textrm{tr} (\hat Z)\vert^2 \right) \nonumber \\
& = 1-p
\end{flalign}
The channel fidelity also provides a means of optimizing the encoding and decoding operations shown in Figure \ref{fig:generic} based on `semi-definite programming' (SDP), a well-studied class of optimization problems \cite{PhysRevLett.94.080501}. 

The channel fidelity of the composite channel (referred to as $\Gamma$, see Equation \ref{eq:channelfidelity}) can be written as a vector inner product by stacking the columns of the matrices $\mathcal{D} \otimes \hat 1 [\ket{\Omega}\bra{\Omega}]$ and $\left( \mathcal{T} \circ \mathcal{E} \circ \mathcal{M} \circ \mathcal{A}\right)_{\ast} \otimes \hat 1 [\ket{\Omega}\bra{\Omega}]$, where $\Lambda_{\ast}$ is the Schr\"odinger dual of a channel ($\Lambda_{\ast}[\rho] = \sum_j \Lambda_j^{\dagger} \rho \Lambda_j$, cf. Equation \ref{eq:sumoperator}). The CPTP constraint $\sum_k \mathcal{D}_k^{\dagger}\mathcal{D}_k = \hat 1$ can be expressed in SDP. If we do not apply additional constraints, the optimal encoder found by this procedure incorporates the following CPTP channel:
\begin{flalign}
& \dfrac{1}{2}\left \lbrace \hat 1 \otimes \ket{00}\bra{00},\,\hat 1 \otimes \ket{00}\bra{01}, \right. \nonumber \\ 
& \left. \hat 1 \otimes \ket{00}\bra{10},\,\hat 1 \otimes \ket{00}\bra{11}\right \rbrace
\end{flalign}
This channel, when implemented before the encoding unitary, returns the ancilla to the state $\ket{00}$ with 100\% fidelity. Physically, this is equivalent to allowing the ancilla to come to equilibrium with a heat bath at zero temperature \cite{mikeandike}, and it cannot be implemented using a unitary operator. In order to find an optimal unitary, we further restrict the SDP to unital channels (those that obey the constraint $\sum_k \mathcal{D}_k\mathcal{D}_k^{\dagger} = \hat 1$). Upon finding the channel fidelity of the optimal channel subject to these constraints, a unitary operator which replicates this fidelity is found by inspection. According to the scheme in \cite{reimpellthesis}, SDP can be iterated to find a code which has a globally optimized encoder, and a globally optimized decoder, and is, on the whole, at least locally optimal.
\subsection{Model Channel}
In contrast with the ideal situation typically studied, each of the ancilla qubits in an NMR experiment is prepared in a mixed state. This is equivalent to the action of a second error channel which occurs before the encoding operation. A mixing channel which works on all inputs is identical to the depolarization channel \cite{mikeandike}; a simpler channel can be used when the input state is $\ket{0}$:
\begin{equation}
M_0 = \sqrt{1-\frac{q}{2}}\hat 1; \quad M_1 = \sqrt{\frac{q}{2}} \hat X;\quad \mathcal{M}=\left \lbrace M_0,\,M_1\right \rbrace.
\label{eq:mixingchannel}
\end{equation}
Treating the presence of entropy in the ancilla qubits as the result of a CPTP channel acting on the state $\ket{0}$ simplifies the calculation of channel fidelities for the error correction process. 
\subsection{Results}
The code which maximizes $F_C$ is presented in Figure \ref{fig:tofcode}:
\begin{figure}[h!]
\centering
\includegraphics[width=0.35\textwidth]{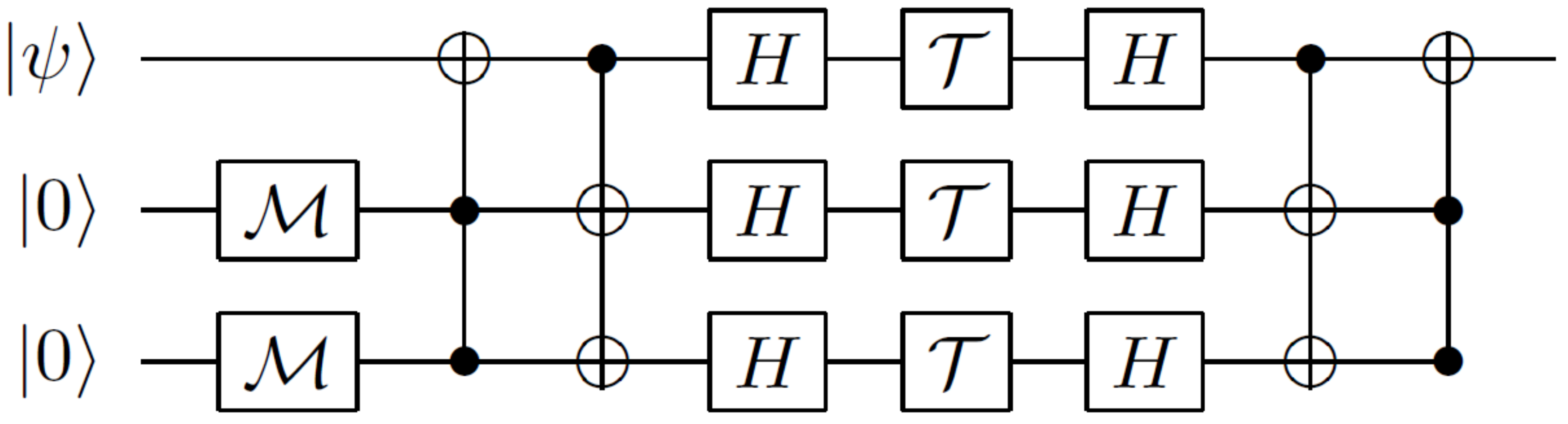}
\caption{Optimal 3-qubit error correcting code against dephasing, when the ancilla qubits used are in mixed states.}
\label{fig:tofcode}
\end{figure}

A simple argument illustrates the function of the added Toffoli gate in this code. Assuming that the message qubit begins in the pure state $\ket{\psi}$, and that no error occurs while the encoded state is being transmitted. Without the added Toffoli gate, the unmodified three-bit code falsely detects an error on the message qubit and flips its state to $\hat X \ket{\psi}$. With the added Toffoli gate, the message qubit is flipped before it is encoded, and the error-correcting gate at the end restores the original state $\ket{\psi}$. 

\begin{figure}
\centering
\includegraphics[width=0.45\textwidth]{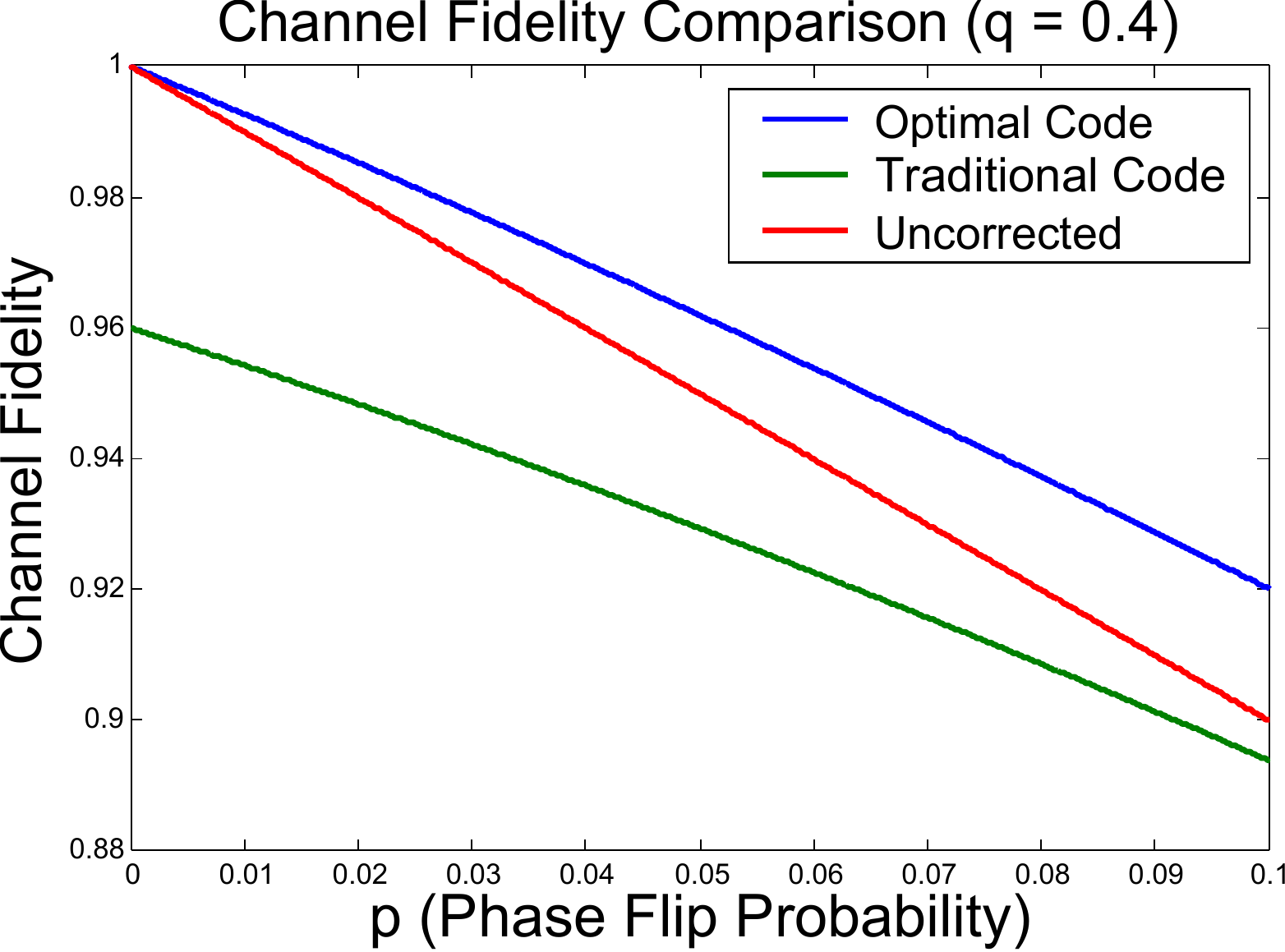}
\caption{The channel fidelity of three processes: transmission through a bit-flip channel, error correction with the traditional code and error correction with the code discussed above. Unlike the traditional code acting on clean qubits, the leading-order error term remains linear, although the $q^2$ term is eliminated.}
\label{fig:fidcurves}
\end{figure}

The fidelities of both the traditional three-bit code and the optimal code can be expressed in terms of $q$, the mixing parameter introduced in Equation \ref{eq:mixingchannel}:
\begin{multline}
F_{C(\textrm{Traditional})}=\left(1-\frac{q^2}{4}\right) - \left( 2q-\dfrac{3q^2}{2}\right)p \nonumber \\-\left( 3-6q+3q^2\right)p^2 +\left(2-4q+2q^2\right)p^3 
\end{multline}
\begin{multline}
F_{C\left(\textrm{Optimal}\right)}=1 - \left( 2q-\dfrac{q^2}{2}\right)p \nonumber \\-\left( 3-6q+3q^2\right)p^2 +\left(2-4q+q^2\right)p^3 \vspace{-24pt}
\end{multline}
If both ancilla qubits have a polarization 0.6 ($q=1-\epsilon = 0.4$, corresponding to the electron polarization at approximately 2 K), the difference between these fidelities is readily apparent; it is seen in Figure \ref{fig:fidcurves}.

Fidelities for these codes can be expressed as functions of the ancilla density matrix elements. Thermalization is assumed to have occurred before the error correction procedure begins, producing a diagonal ancilla density matrix:
\begin{flalign}
\rho_{\mathrm{anc}} &= \rho_{00}\ket{00}\bra{00} +\rho_{01} \ket{01}\bra{01} \nonumber \\& \quad +\rho_{10} \ket{10}\bra{10}+\rho_{11} \ket{11}\bra{11}
\end{flalign}
\begin{flalign}
\textrm{for which }F_{C\left(\mathrm{Traditional}\right)} = \rho_{11}(2p-1)+ \nonumber \\(1-p)(1+(p-2p^2)(1-2(\rho_{10}+\rho_{01}))),\\
F_{C\left(\mathrm{Optimal}\right)} = (1-p)(1+(p-2p^2)(2\rho_{00}-1)).
\end{flalign}

\begin{figure}
\centering
\includegraphics[width=0.45\textwidth]{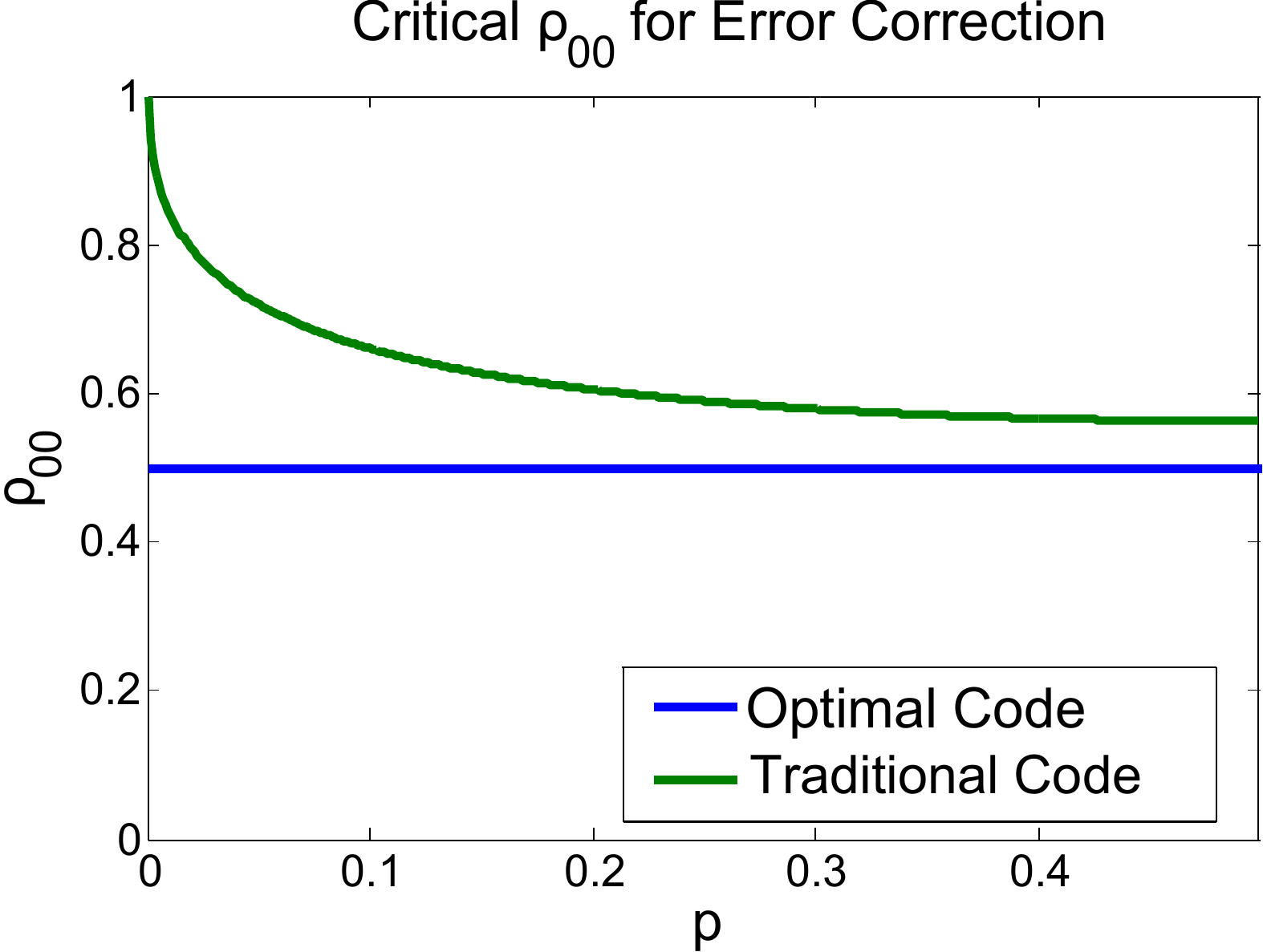}
\caption{Ancilla density matrix element required in order for error correction to improve the fidelity of a state exposed to a noisy channel. This is the criterion which determines whether or not the optimal error correction code produces a fidelity which is greater than $1-p$, the fidelity produced by inaction. Since the fidelity produced by the traditional code does not depend exclusively on $\rho_{00}$, the matrix element is calculated from $q$, the mixing parameter.}
\label{fig:rhooneone}
\end{figure}

$F_{C\left(\mathrm{Optimal}\right)}$ is greater than $1-p$ whenever $\rho_{00}>\nicefrac{1}{2}$. This required $\rho_{00}$ is contrasted with the critical density matrix element for correction using the traditional code in Figure \ref{fig:rhooneone}. 
$\rho_{00}>\nicefrac{1}{2}$ will serve as the criterion for useful error correction. However, the equilibrium room-temperature polarization of carbon-13 (a commonly-used nucleus) in a 7 T field results in $\rho_{00}=\nicefrac{1}{4}+4.8 \times 10^{-7}$. If refrigeration were used to enhance the polarization of these qubits, the temperature required would be 0.6 mK. It is impossible to adequately polarize these qubits with current (or near-future) technology by increasing the magnetic field, since
\begin{flalign}
\epsilon = \tanh \left( \dfrac{\hbar \gamma B_0}{k_B T}\right), \,\, \therefore \,\, T_{\epsilon = k} \propto B_0.
\end{flalign}
Doubling the magnetic field present (a technical challenge) raises the temperature for constant polarization to 1.2 mK. A different technique for enhancing the polarization of a quantum register is, therefore, necessary.
\section{Heat Bath Algorithmic Cooling}
\subsection{Background}
Given the need for a different cooling technique, Schulman and Vazirani \cite{Umesh99molecularscale} outlined a procedure to extract a small number of highly-polarized qubits from a large ensemble with low polarization. Using the closed-system algorithm presented there, it is possible to extract $O\left(\epsilon^2 n\right)$ qubits with a constant bias from a register of $n$ qubits with initial polarization $\epsilon$. To extract the required three qubits with polarization $\sqrt{2}-1$ (which is required in order to satisfy the requirement that $\rho_{00}>\nicefrac{1}{2}$) from a register with an initial polarization of $\sim 10^{-6}$ would require a molecule with $\sim 6 \times 10^{13}$ spins, clearly an impractical criterion. 

Open system methods provide a much better prospect for implementation. An efficient method has already been implemented at the date of this writing, the partner-pairing algorithm introduced by Schulman et al \cite{PhysRevLett.94.120501} and implemented by Ryan et al. \cite{PhysRevLett.100.140501}.

\subsection{The Partner Pairing Algorithm}
The partner pairing algorithm was first discussed in \cite{PhysRevLett.94.120501}, where it was found to be optimally efficient (it provides the maximum polarization increase possible per operation) and scalable (the number of elementary operations scales polynomially with the system size). There are two elementary operations in the partner pairing algorithm; compression and exchange. Compression applies a unitary operator to the density matrix of a thermally-mixed state, arranging the density matrix elements so that the element corresponding to $\ket{0}^{\otimes n}\bra{0}^{\otimes n}$ is the largest element, and the rest are non-increasing. Exchange replaces the last qubit in the thermally-mixed register with one from the heat bath; it is the net effect of allowing that qubit to come to thermal equilibrium with the bath. 

These two operations, combined and iterated, will bias the density matrix of the ancilla toward $\ket{0}^{\otimes n}$. Since the states involved are classical, and the operations are simple, the effect of this algorithm is calculated numerically in \textsc{matlab}. There are two situations in which  this technique will be used; initial state preparation and ancilla refreshing. It is important not to cool the qubit which stores the information to be preserved by error correction while refreshing the ancilla. Also, note that the partner pairing algorithm requires three qubits in order to obtain polarization beyond that of the bath, because once two qubits are polarized to the same $\tilde \epsilon_{\textrm{bath}}$, their density matrix is of the form
\begin{equation*}
\rho = \left( \begin{array}{cccc}
\left(1+\tilde{\epsilon}_{\textrm{bath}}\right)^2 & 0 & 0 & 0 \\
0 & 1-\tilde{\epsilon}_{\textrm{bath}}^2 & 0 & 0 \\
0 & 0 & 1-\tilde{\epsilon}_{\textrm{bath}}^2 & 0 \\
0 & 0 & 0 & \left(1-\tilde{\epsilon}_{\textrm{bath}}\right)^2
\end{array} \right).
\end{equation*}
Here, the diagonal density matrix elements are already ordered, so there is no way to perform a compression step. The temperature required for a bath polarization needed to perform useful error correction ($\rho_{00}>\nicefrac{1}{2}$) is then 3.4 Kelvin, lower than what can be achieved by immersing the NMR sample in liquid helium. Therefore, it is necessary for experimental feasibility to introduce a fourth qubit to be used to refresh ancilla qubits. This qubit can also be used to enhance the results of state preparation. Since the compression step of each HBAC iteration leaves the qubits in the register being cooled at different polarizations, we choose to use the most polarized qubits in the error-correction ancilla, and reserve the most polarized remaining qubit for the message to be transmitted. The purity of this qubit is shown in Figure \ref{fig:polconmap2}.
\begin{figure}[ht]
\centering
\includegraphics[width=0.4\textwidth]{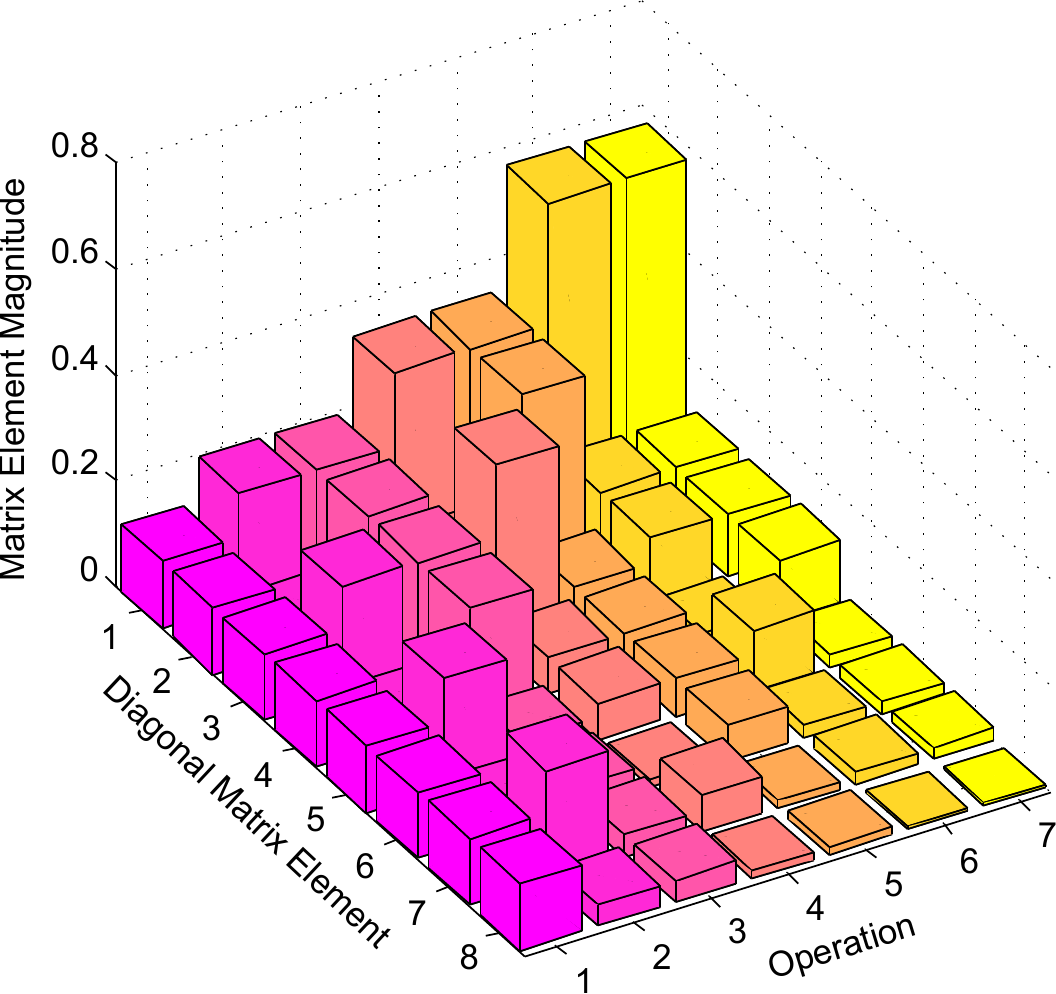}
\caption{The effect of the partner pairing algorithm on the diagonal elements of an initially highly mixed thermal state. The even-numbered operations are exchanges between the register and the polarization bath, and the odd-numbered operations are unitary operations which sort the density matrix elements according to size.}
\label{fig:hbacbarplot}
\end{figure}
\subsection{Results}
In this and the following section, we describe the physical parameters required to perform multiple rounds of error correction in an experiment and assess the feasibility of such an experiment.

\begin{figure}
\centering
\includegraphics[width=0.45\textwidth]{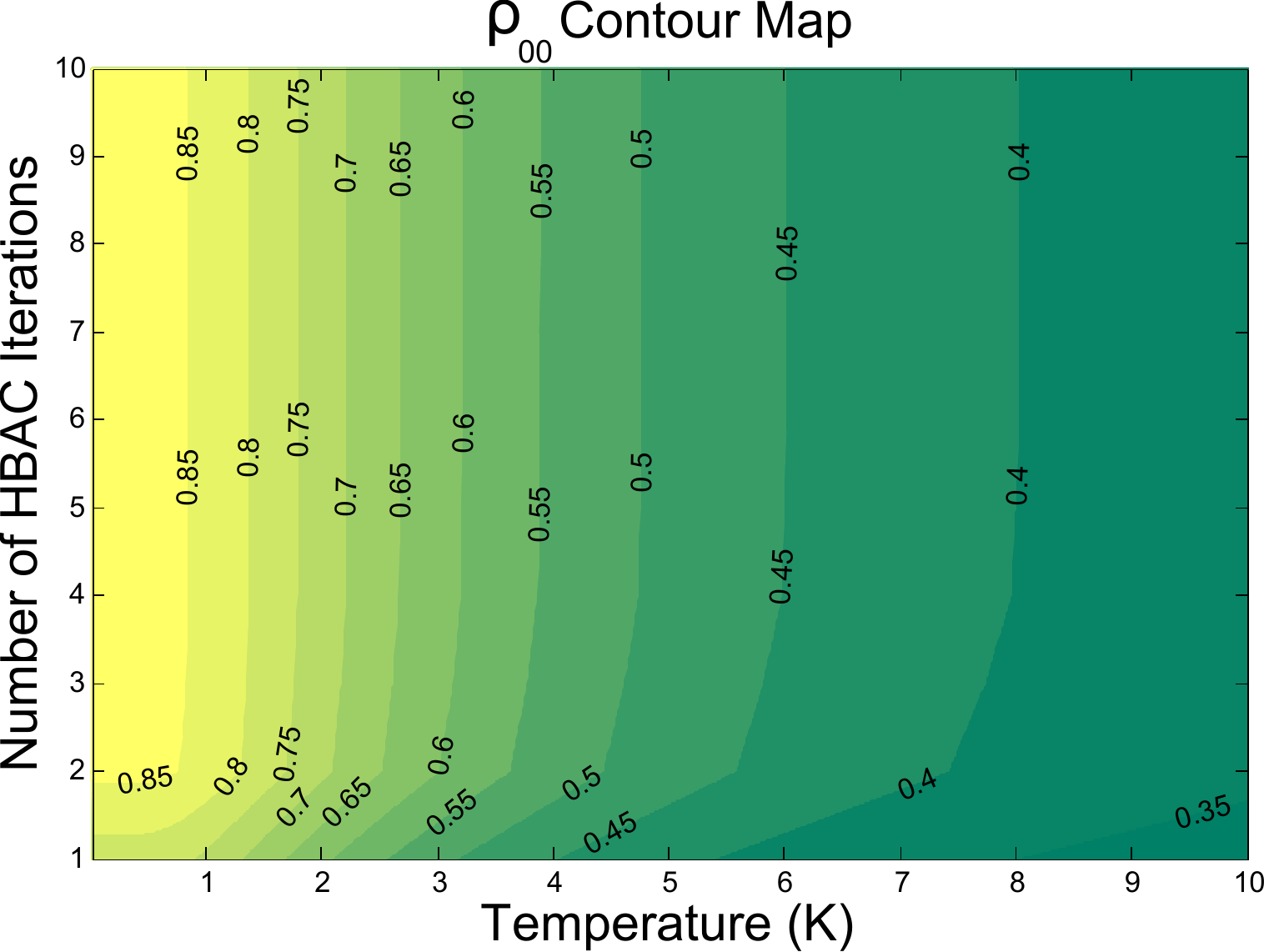}
\caption{The ancilla density matrix element after refreshing between error correction steps. A value above 0.5 indicates that useful $(F_C > 1-p)$ error correction can take place.}
\label{fig:ancillapol}
\end{figure}

\begin{figure}
\centering
\includegraphics[width=0.45\textwidth]{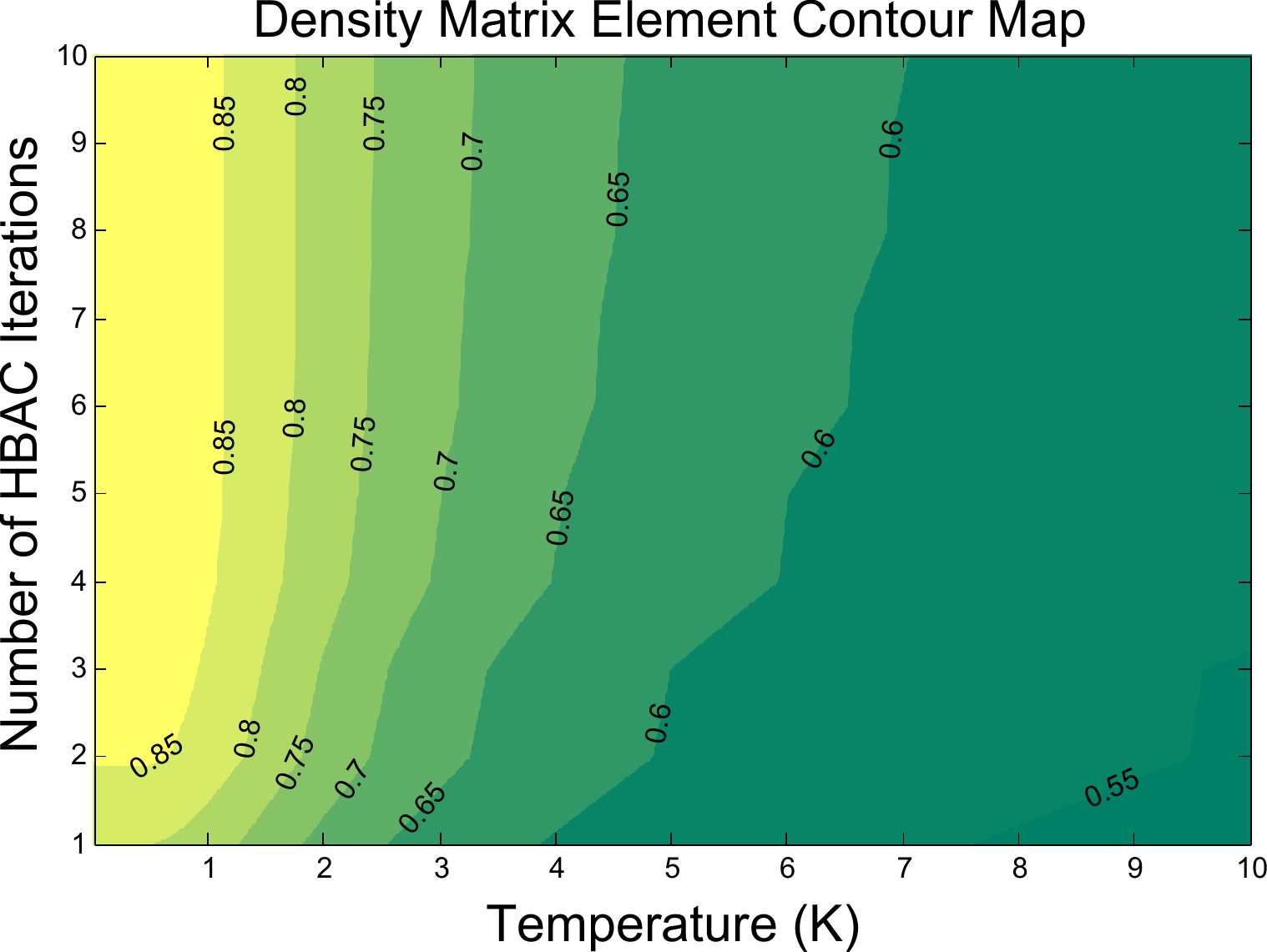}
\caption{The density matrix element corresponding to $\ket{0}\bra{0}$ on the message qubit immediately following initialization.}
\label{fig:polconmap2}
\end{figure}

The most important use of algorithmic cooling in this proposal is to refresh ancilla qubits between rounds of error correction. (Algorithmic cooling will also be used to prepare the initial states used in error correction, but refreshing is the critical step for performing multiple iterations.) In order to obtain the $\rho_{00}$ required for error correction, we subject three qubits to the partner-pairing algorithm, and use the two most polarized for the ancilla, assuming that the remaining qubit will be thermalized before the next refreshing stage. (This qubit is still necessary to facilitate polarization of the other two, as described in the previous section). We present in Figure \ref{fig:ancillapol} the achievable $\rho_{00}$ as a function of the temperature and number of iterations of the partner-pairing algorithm to be performed on the register. From this, we see that temperatures above 4.7 K will not produce usable ancilla qubits. Any temperature below this will be useful, with colder temperatures being preferable. Also, we see that there is very little beneficial effect to performing more than three iterations of algorithmic cooling, as long as the temperature is below 4.7 K. Since the aim of algorithmic cooling is to produce polarization on two qubits, initialization with the same number of iterations and the same temperature will result in a third qubit with considerable polarization. This polarization is displayed in a similar contour map in Figure \ref{fig:polconmap2}. This polarization is uncommonly large for NMR experiments, increasing the efficacy of the protocol as compared to a typical NMR experiment.

\begin{figure*}[h!]
\centering
\includegraphics[width=0.85\textwidth]{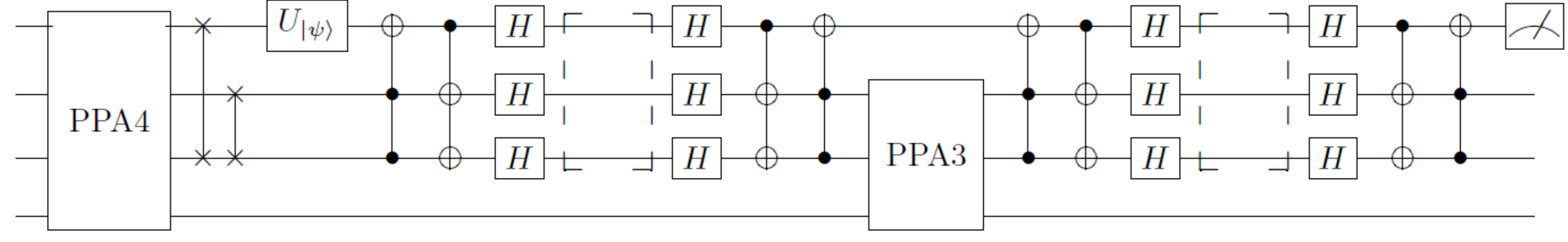}
\caption{Proof of concept/testbed algorithm for multiple rounds of error correction. After the partner-pairing algorithm is used to initialize the register, an arbitrary 1-qubit state is prepared on the qubit with the third-highest polarization, the two most polarized qubits being used as an ancilla. Within the dotted rectangles, fault-tolerant computations can be implemented. Note that the message qubit is left unencoded while the ancilla is being refreshed. The fourth qubit is employed only in the partner-pairing algorithm in order to obtain polarization in excess of the bath polarization. Since the number of iterations in the partner-pairing algorithm is variable, its instances in the algorithm above have been left as subroutines.}
\label{fig:testcode}
\end{figure*}

\begin{figure*}[h!]
\centering
\includegraphics[width=0.85\textwidth]{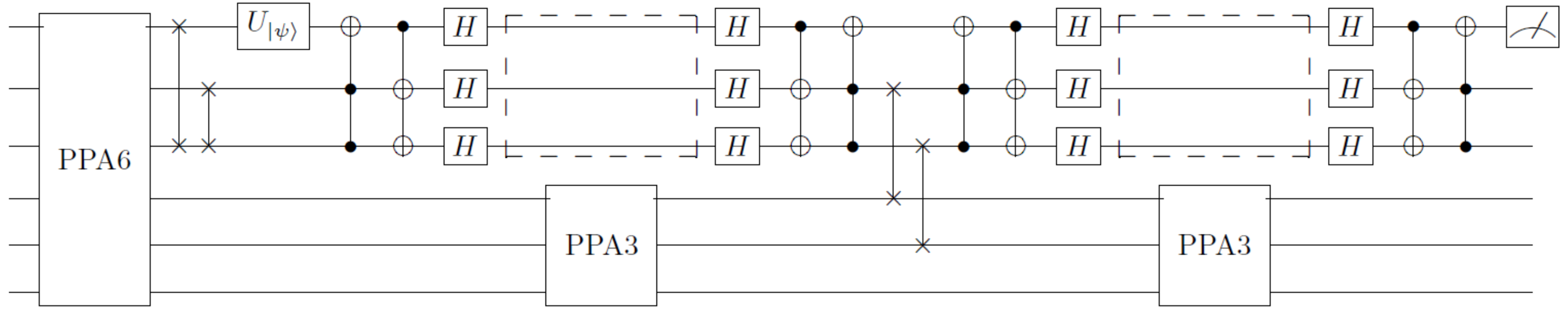}
\caption{Plausible algorithm for storage and computation on one logical qubit. After the partner-pairing algorithm is used to initialize the register, an arbitrary 1-qubit state is prepared. Here, an extra two-qubit register is used for error correction while the other 2-qubit ancilla is being refreshed. If, in the experiment chosen, the partner pairing algorithm can be implemented on non-adjacent qubits (as is the case in solid state NMR), then the two swap gates between decoding and encoding can be eliminated. The message qubit can then be encoded at all times, and the chief source of error is imperfect control.}
\label{fig:realcode}
\end{figure*}

It is important to note that the number of qubits which are readily available for experiments in NMR is limited. With this in mind, two algorithms are proposed; one which contains all of the necessary stages to execute multiple rounds of error correction with only four qubits, and one which has six qubits, so that the message state need not be unencoded while ancilla qubits are being refreshed. The four-qubit code recommended for experimental implementation is presented in Figure \ref{fig:testcode}. The six-qubit code recommended for practical extraction of entropy is presented in Figure \ref{fig:realcode}.
We determine that, given perfect control of the unitary operations we implement, that three iterations of HBAC will be required (the number implemented in \cite{PhysRevLett.100.140501}) at a temperature below 4.7 K, which can be achieved in NMR. Coherent control over electron degrees of freedom will also be required, which is experimentally feasible. In the next section, we re-assess the feasibility of an NMR experiment to perform multiple rounds of error correction, given the fact that implementation of unitary operators in experiment introduces quantifiable noise. 
\subsection{Effects of Imperfect Control}
Since the goal of error correction is to reduce the effects of decoherence, it is important to examine the decoherent effects of imperfect implementation of the unitary operations used for error correction. In \cite{RLL09}, it is found that the average error per gate implemented on malonic acid is $\sim 1\%$. This figure can be used to further determine the utility of the optimal error-correcting code, and the exterior temperature required for continual iteration of the error-correction and refreshing procedures. 

\begin{figure}[h!]
\centering
\includegraphics[width=0.45\textwidth]{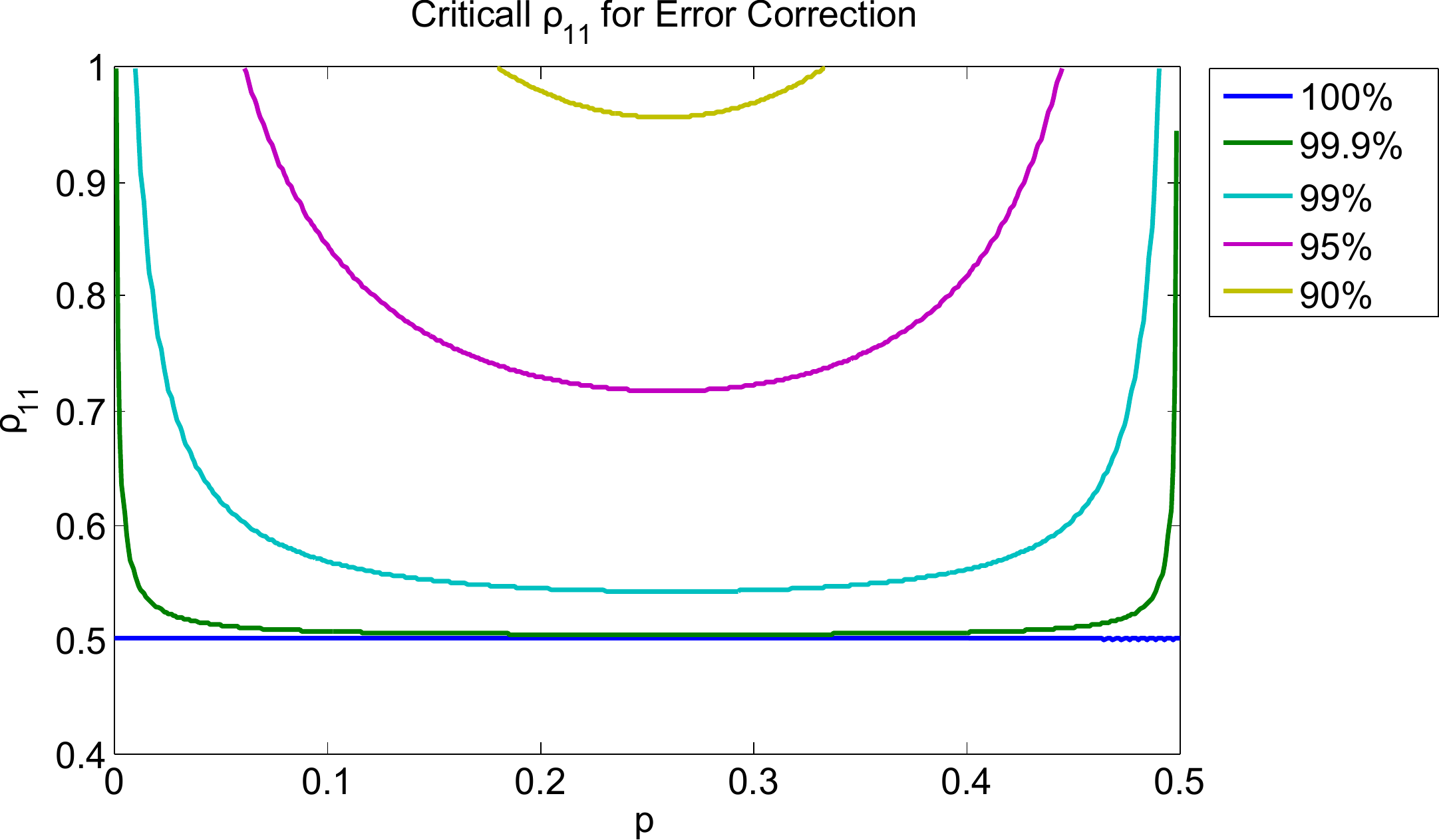}
\caption{Ancilla density matrix element required to perform useful error correction, analogous to Figure \ref{fig:rhooneone}. Curves based on gate fidelities given in the legend are shown. Note the presence of the curve with $99\%$ gate fidelity, this results from the estimated $1\%$ error rate for single-qubit gates in NMR.}
\label{fig:rainbow}
\end{figure}

First, we examine the effect of imperfect control on the error correction unitaries; encoding and decoding the fault-tolerant state. In order to model imperfect unitaries, we create a channel consisting of the desired unitary, followed by depolarization \cite{mikeandike} on all three qubits with error parameter $c$.
\begin{equation}
\tilde{U} =\left \lbrace \sqrt{1-\dfrac{3c}{4}}\hat 1,\,\,\sqrt{\dfrac{c}{4}}\hat X,\,\,\sqrt{\dfrac{c}{4}} \hat Y,\,\,\sqrt{\dfrac{c}{4}}\hat Z \right \rbrace^{\otimes 3} U. \label{eq:dep}
\end{equation}
We represent the efficacy of a given quantum operation using the channel fidelity of the depolarizing map on three qubits:
\begin{flalign}
F\left( U,\,\,\tilde{U} \right)= F_C\left(\left \lbrace \sqrt{1-\dfrac{3c}{4}}\hat 1,\,\,\sqrt{\dfrac{c}{4}}\hat X,\,\,\sqrt{\dfrac{c}{4}} \hat Y,\,\,\sqrt{\dfrac{c}{4}}\hat Z \right \rbrace^{\otimes 3} \right) \nonumber \\
=\left(1 -\dfrac{3c}{4}\right)^3.
\end{flalign}
We plot in Figure \ref{fig:rainbow} the density matrix element required to perform useful error correction at a series of estimated gate fidelities.
 
\begin{figure}
\centering
\includegraphics[width=0.45\textwidth]{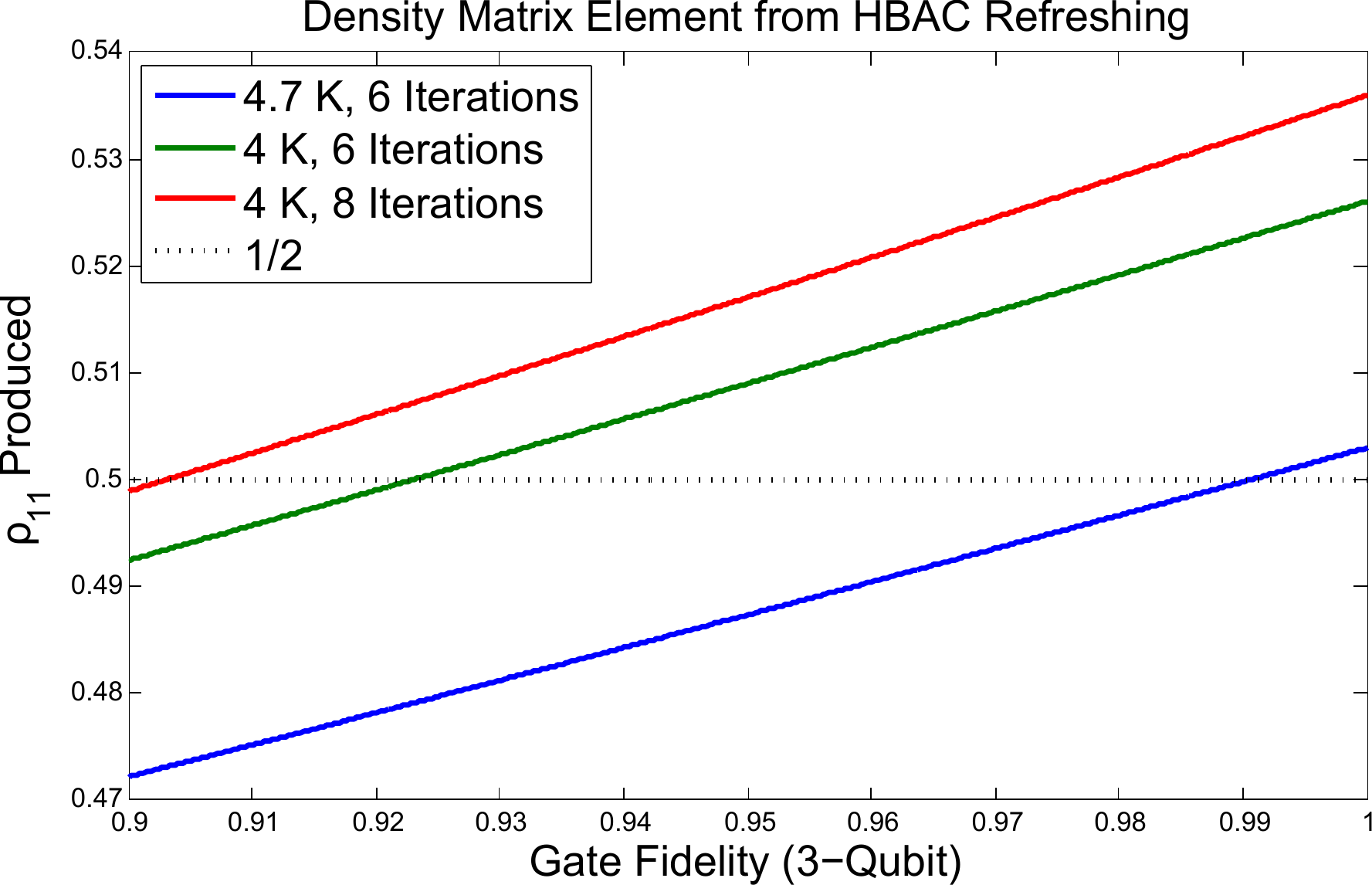}
\caption{Polarization produced by 6 iterations of imperfect HBAC on 3 qubits, with an exterior bath polarization of 0.3.}
\label{fig:robustHBAC}
\end{figure}

\begin{figure}
\centering
\includegraphics[width=0.45\textwidth]{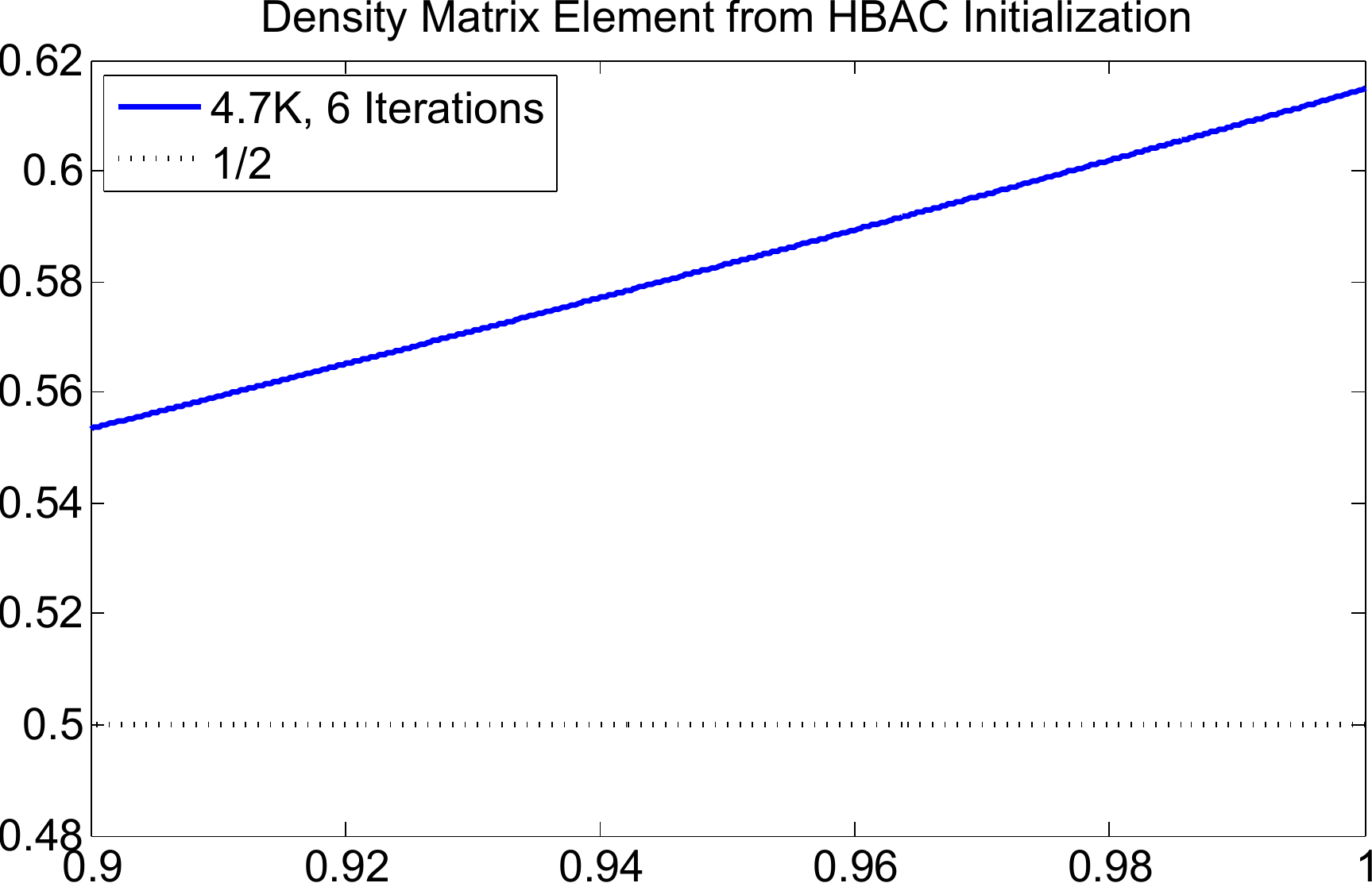}
\caption{Polarization produced by 6 iterations of imperfect HBAC on 4 qubits, with an exterior bath polarization of 0.3.}
\label{fig:robustHBAC2}
\end{figure}

We now examine the impact of imperfect control on algorithmic cooling itself. We implement a depolarizing map after each compression operation, similar to the process presented in Equation \ref{eq:dep}. In Figure \ref{fig:robustHBAC}, we present the results of refreshing with imperfect control for 6 and 8 iterations of HBAC, with a bath polarization of 0.31 and 0.36 (corresponding to temperatures of 4.7 and 4 K, respectively). In Figure \ref{fig:robustHBAC2}, the same analysis is undertaken for initialization. 
\section{Conclusion}
Here, we have analysed the use of error correction for data storage and processing. The transport of entropy from a qubit to be preserved to the exterior environment is proposed, as is the use of an intermediate heat bath. The procedures outlined here for state initialization and ancilla re-polarization are experimentally achievable with existing ESR/NMR equipment, as well as in other potential quantum computing architectures where a heat bath is present. 
\section{Acknowledgements}
Ben Criger thanks the Advanced Optimization Lab at McMaster University for continuing the development of SeDuMi, the SDP solving package employed in this manuscript. The authors also thank NSERC, CIFAR, and QuantumWorks for their continued support. IQC thanks Industry Canada for their continued support.
\end{document}